\documentclass{llncs}

\usepackage{amssymb,amsmath,stmaryrd}
\usepackage{latexsym}
\usepackage{amsfonts}
\usepackage{layout}
\usepackage{picinpar}
\usepackage{paralist}
\usepackage{times}

\usepackage[ruled, vlined]{algorithm2e}
\dontprintsemicolon

\usepackage{epic}
\usepackage{eepic}
\usepackage{graphicx}
\usepackage{comment}
\usepackage{gastex}
\usepackage{url}
\usepackage{wrapfig}

\newcommand{\abs}[1]{\ensuremath{\lvert #1\rvert}}
\newcommand{\tuple}[1]{\langle #1 \rangle}

\renewcommand{\L}{\mathsf{L}}
\newcommand{\R}{\mathsf{R}}
\newcommand{\Last}{{\sf Last}}
\newcommand{\rank}{{\sf rank}}
\newcommand{\weight}{w}
\newcommand{\Mem}{{\sf Mem}}
\newcommand{\obciach}{\upharpoonright}
\newcommand{\nats}{\mathbb{N}}

\newcommand{\va}{{\mathsf Val}}
\newcommand{\cale}{{\mathcal E}}
\newcommand{\calw}{{\mathcal W}}

\newcommand{\nat}{\mathbb N} 
\newcommand{\zed}{{\mathbb Z}}

\newcommand{\set}[1]{\{#1\}}

\newcommand{\rat}{{\mathbb Q}}

\newcommand{\real}{{\mathbb R}}

\newcommand{\trans}{{\delta}}
\newcommand{\dist}{{\cal D}}
\newcommand{\distr}{\dist}
\newcommand{\Supp}{{\sf Supp}}

\newcommand{\straa}{\sigma}
\newcommand{\Straa}{\Sigma}

\newcommand{\pat}{\omega}
\newcommand{\Paths}{\Omega}
\newcommand{\Aa}{{\cal A}}
\newcommand{\Prb}{\mathbb{P}}
\newcommand{\Exp}{\mathbb{E}}

\newcommand{\Inf}{{\sf Inf}}
\newcommand{\EL}{{\sf EL}}
\newcommand{\MP}{{\sf MP}}
\newcommand{\Parity}{{\sf Parity}}
\newcommand{\PosEnergy}{{\sf PosEnergy}}
\newcommand{\MeanPayoff}{{\sf MeanPayoff}}

\newcommand{\Win}{{\sf Win}}

\newcommand{\ov}{\overline}
\newcommand{\Reach}{{\sf Reach}}
\newcommand{\vaMP}{{\sf ValMP}}

\makeatletter
\DeclareRobustCommand\sfrac[1]{\@ifnextchar/{\@sfrac{#1}}%
                                            {\@sfrac{#1}/}}
\def\@sfrac#1/#2{\leavevmode\scalebox{.9}{\kern.1em\raise.5ex
         \hbox{$\m@th\mbox{\fontsize\sf@size\z@
                           \selectfont#1}$}\kern-.1em
         /\kern-.15em\lower.25ex
          \hbox{$\m@th\mbox{\fontsize\sf@size\z@
                            \selectfont#2}$}}}
\DeclareRobustCommand\numfrac[1]{\@ifnextchar/{\@numfrac{#1}}%
                                            {\@numfrac{#1}}}
\def\@numfrac#1{\leavevmode \hbox{$\m@th\mbox{\fontsize\sf@size\z@
                           \selectfont#1}$}}
\makeatother

\newcommand{\sink}{{\sf sink}}

\title{
Energy and Mean-Payoff Parity \\ Markov Decision Processes
}

\author{Krishnendu Chatterjee\inst{1} \and Laurent Doyen\inst{2}}

\institute{
IST Austria (Institute of Science and Technology Austria) \\
\and LSV, ENS Cachan \& CNRS, France 
}

\begin{document}
\maketitle

\begin{abstract} 
We consider Markov Decision Processes (MDPs) with mean-payoff parity and 
energy parity objectives.
In system design, the parity objective is used to encode $\omega$-regular 
specifications, and the mean-payoff and energy objectives can be used to model 
quantitative resource constraints. 
The energy condition requires that the resource level never drops below~$0$, 
and the mean-payoff condition requires that the limit-average value of the resource 
consumption is within a threshold. 
While these two (energy and mean-payoff) classical conditions are equivalent 
for two-player games, we show that they differ for MDPs. 
We show that the problem of deciding whether a state is almost-sure winning 
(i.e., winning with probability~$1$) in energy parity MDPs is in NP~$\cap$~coNP, 
while for mean-payoff parity MDPs, 
the problem is solvable in polynomial time, improving a recent PSPACE bound.
\end{abstract}

\section{Introduction}

Markov decision processes (MDPs) are a standard model for 
systems that exhibit both stochastic and nondeterministic 
behaviour. The nondeterminism represents the freedom
of choice of control actions, while the probabilities
describe the uncertainty in the response of the system to control actions.
The control problem for MDPs asks whether
there exists a strategy (or policy) to select control
actions in order to achieve a certain goal with a certain
probability. 
MDPs have been used in several areas
such as planning, probabilistic reactive programs,  
verification and synthesis of (concurrent) probabilistic
systems~\cite{CY95,Var85,BdA95}.

The control problem may specify a goal as a set of desired
traces (such as $\omega$-regular specifications), or 
as a quantitative optimization objective for a payoff function on 
the traces of the MDP. Typically, discounted-payoff and
mean-payoff functions have been studied~\cite{FV97}.
Recently, the energy objectives (corresponding to total-payoff functions)
have been considered in the design of resource-constrained 
embedded systems~\cite{BFLMS08,talk-pacuk,CD10a} such as power-limited systems, 
as well as in queueing processes, and gambling models (see also~\cite{soda10} and references therein).
The energy objective
requires that the sum of the rewards be always nonnegative along a trace.
Energy objective can be expressed in the setting of boundaryless one-counter MDPs~\cite{soda10}.
In the case of MDPs, achieving energy objective with probability~$1$
is equivalent to achieving energy objective in the stronger setting 
of a two-player game where the probabilistic choices are replaced by adversarial choice. 
This is because if a trace~$\rho$ violates the energy condition in the game,
then a finite prefix of~$\rho$ would have a negative energy, and this finite prefix has positive
probability in the MDP. Note that in the case of two-player games, the energy
objective is equivalent to enforce nonnegative mean-payoff value~\cite{BFLMS08,BCDGR10}. 

In this paper, we consider MDPs equipped with the combination of a parity objective
(which is a canonical way to express the $\omega$-regular conditions~\cite{Thomas97}),
and a quantitative objective specified as either mean-payoff or energy condition.
Special cases of the parity objective include reachability and fairness objectives
such as B\"uchi and coB\"uchi conditions. Such combination of quantitative 
and qualitative objectives is crucial in the design of reactive systems with 
both resource constraints and functional requirements~\cite{CAHS03,ChatterjeeHJ05,BFLMS08,BCHJ09}.
For example, Kucera and Strazvosky consider the combination of PCTL with
mean-payoff objectives for MDPs and present an EXPTIME algorithm~\cite{KucStr05}.
In the case of energy parity condition, it can also be viewed as a natural 
extension of boundaryless one-counter MDPs with fairness conditions.

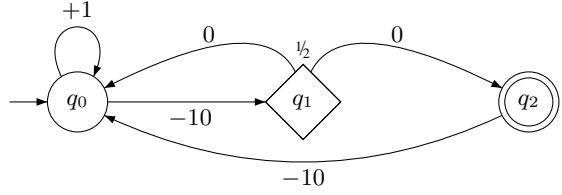
\begin{figure}[!tb]
  \begin{center}
    \hrule
   \begin{picture}(75,28)(0,0)


\node[Nmarks=i, Nframe=y](p1)(10,13){$q_0$}

\rpnode[Nmarks=n](p2)(40,13)(4,5){$q_1$}
\rpnode[Nmarks=n, ExtNL=y, NLangle=92, NLdist=1](p2)(40,13)(4,5){$\sfrac{1}{2}$}
\node[Nmarks=r, Nframe=y, rdist=0.8](p3)(70,13){$q_2$}   

\drawbpedge[ELpos=50, ELside=r, ELdist=.5](p2,95,15,p1,28,10){$0$}
\drawbpedge[ELpos=50, ELside=l, ELdist=.5](p2,85,15,p3,152,10){$0$}

\drawloop[ELpos=50, ELside=l,loopCW=y, loopangle=90, loopdiam=6](p1){$+1$}
\drawedge[ELpos=50, ELside=r, curvedepth=0](p1,p2){$-10$}
\drawedge[ELpos=50, ELside=l, curvedepth=8](p3,p1){$-10$}


\end{picture}
    \hrule
    \caption{An energy B\"uchi MDP. The player-$1$ states are $q_0,q_2$, and
the probabilistic state is $q_1$. \label{fig:energy-buchi-MDP}}
  \end{center}
\vspace{-2em}
\end{figure}

Consider the MDP in \figurename~\ref{fig:energy-buchi-MDP}, with the objective
to visit the B\"uchi state~$q_2$ infinitely often, while maintaining the energy 
level (i.e., the sum of the transition weights) nonnegative. A winning strategy 
from~$q_0$ would loop $20$ times on~$q_0$ to accumulate energy and then it can afford 
to reach the probabilistic state from which the B\"uchi state is reached with
probability$\,\sfrac{1}{2}$ and cost~$20$. If the B\"uchi state is not reached
immediately, then the strategy needs to recharge $10$ units of energy and try again.
This strategy uses memory and it is also winning with probability~$1$ 
for the nonnegative mean-payoff B\"uchi objective. 
In general however, the energy and mean-payoff parity objectives do not coincide (see later
the example in~\figurename~\ref{fig:energy-vs-mean-payoff}). In particular,
the memory requirement for energy parity objective is finite (at most exponential)
while it may be infinite for mean-payoff parity.


We study the computational complexity of the problem of deciding if there exists
a strategy to achieve energy parity objective, or mean-payoff parity objective with 
probability~$1$ (i.e., almost-surely). 
We provide tight bounds for this problems in the following sense.
\begin{compactenum}
\item For energy parity MDPs, we show that the problem is in NP~$\cap$~coNP, and present
a pseudo-polynomial time algorithm.
Our bounds are the best conceivable upper bound unless parity games can be solved in P\footnote{Parity games  
polynomially reduce to two-player energy games~\cite{Jurdzinski98,BFLMS08,BCDGR10}, 
and thus to energy MDPs. Hence the problem for 
almost-sure energy parity MDPs is at least as hard as solving two player parity games.}, which 
is a long-standing open question. 

\item For mean-payoff parity MDPs, we show that the problem 
is solvable in polynomial time (and thus PTIME-complete).
Our result improves the recent PSPACE upper 
bound of~\cite{GIMBERT:2011:HAL-00559173:2} for this problem. 
\end{compactenum}
We refer to~\cite{CY95,GH10,CH11} 
for importance of the computation of almost-sure winning set related to 
robust solutions (independence of precise transition probabilities) and the more general 
quantitative problem. 
The computation of the almost-sure winning set in MDPs typically relies either on the 
end-component analysis, or analysis of attractors and sub-MDPs. 
The result of~\cite{GIMBERT:2011:HAL-00559173:2} for mean-payoff parity MDPs
uses the analysis of attractors and sub-MDPs and gives a nested recursive PSPACE algorithm 
for the problem similar to the classical algorithm for parity games. 
Our results rely on the end-component analysis, but in a much more refined 
way than the standard analysis, to obtain a polynomial-time algorithm.
Our proof combines techniques for mean-payoff and parity objectives to 
produce infinite-memory strategy witnesses, which is necessary in general.
We present an algorithm that iterates successively over 
even priorities $2i$ and computes almost-sure winning end-components with the even 
priority $2i$ as the best priority.

For energy parity MDPs the end-component based analysis towards polynomial-time algorithm 
does not work since solving energy parity MDPs is at least as hard as solving 
two-player parity games.
Instead, for energy parity MDPs, we present a quadratic reduction to two-player
energy B\"uchi games which are in NP~$\cap$~coNP and solvable in pseudo-polynomial 
time~\cite{CD10a}. 


Due to lack of space we present the detailed proofs in the appendix.

\section{Definitions}

\paragraph{\bf Probability distributions.}
A \emph{probability distribution} over a finite set $A$ is a function
$\kappa: A \to [0,1]$ such that $\sum_{a \in A} \kappa(a) = 1$. 
The \emph{support} of $\kappa$ is the set $\Supp(\kappa) = \{a \in A \mid \kappa(a) > 0\}$.
We denote by $\dist(A)$ the set of probability distributions on $A$. 

\paragraph{\bf Markov Decision Processes.}
A \emph{Markov Decision Process} (MDP) $M =(Q, E, \trans)$ 
consists of a finite set $Q$ of states 
partitioned into \mbox{\emph{player-$1$}} \emph{states} $Q_1$ and 
\emph{probabilistic states} $Q_P$ (i.e., $Q = Q_1 \cup Q_P$), 
a set $E \subseteq Q \times Q$ of edges such that for all $q \in Q$,
there exists (at least one) $q' \in Q$ such that $(q,q') \in E$,
and a probabilistic transition function $\trans: Q_P \to \distr(Q)$
such that for all $q \in Q_P$ and $q' \in Q$, we have $(q,q') \in E$ 
iff $\trans(q)(q') > 0$. We often write $\trans(q,q')$ for $\trans(q)(q')$. 
For a state $q\in Q$, we denote by $E(q) = \{q' \in Q \mid (q,q') \in E\}$ 
the set of possible successors of $q$.


\smallskip\noindent{\bf End-components and Markov chains.}
A set $U \subseteq Q$ is \emph{$\trans$-closed} if for all $q \in U \cap Q_P$ 
we have $\Supp(\trans(q)) \subseteq U$. The sub-MDP induced by a 
$\trans$-closed set $U$ is $M \obciach U = (U, E \cap (U \times U), \trans)$.
Note that $M \obciach U$ is an MDP if for all $q \in U$ there exists $q' \in U$
such that $(q,q') \in E$.
A \emph{Markov chain} is a special case of MDP where $Q_1 = \emptyset$.
A \emph{closed recurrent set} for a Markov chain is a $\trans$-closed set $U \subseteq Q$ 
which is strongly connected.
End-components in MDPs play a role 
equivalent to closed recurrent sets in Markov chains.  
Given an MDP $M=(Q,E,\trans)$ with partition $(Q_1,Q_P)$, 
a set $U \subseteq Q$ of states is an \emph{end-component}
if $U$ is $\trans$-closed and the sub-MDP $M \obciach U$ is strongly connected~\cite{CY95,deAlfaro97}.
We denote by $\cale(M)$ the set of end-components of an MDP~$M$.

\smallskip\noindent{\bf Plays.}
An MDP can be viewed as the arena of a game played for infinitely
many rounds from a state $q_0 \in Q$ as follows.
If the game is in a player-$1$ state~$q$, then player~$1$ chooses the successor state 
in the set $E(q)$; otherwise the game is in a probabilistic state~$q$, 
and the successor is chosen according to the probability distribution $\trans(q)$.
This game results in a \emph{play} from~$q_0$, i.e., 
an infinite path $\rho = q_0 q_1 \dots$ such that $(q_i,q_{i+1}) \in E$ for all $i \geq 0$. 
The prefix of length $n$ of $\rho$ is denoted by $\rho(n) = q_0 \dots q_n$,
the last state of $\rho(n)$ is $\Last(\rho(n)) = q_n$.
We write $\Paths$ for the set of all plays.


\smallskip\noindent{\bf Strategies.}
A \emph{strategy} (for player~$1$) is a function
$\straa: Q^*Q_1 \to \dist(Q)$ such that for all $\rho \in Q^*$, $q \in Q_1$, and $q' \in Q_P$,
if $\straa(\rho \cdot q)(q') > 0$, then $(q,q') \in E$.
We denote by~$\Straa$ the set of all strategies. 
An \emph{outcome} of $\straa$ from~$q_0$ is a play $q_0 q_1 \dots$ where
$q_{i+1} \in \Supp(\straa(q_0 \dots q_i))$ for all $i \geq 0$ such that $q_i \in Q_1$. 

\smallskip\noindent{\bf Outcomes and measures.}
Once a starting state $q \in Q$ and a strategy $\straa \in \Straa$
are fixed, the outcome
of the game is a random walk $\pat_q^{\straa}$ for which the
probabilities of every \emph{event} $\Aa \subseteq \Paths$, which 
is a measurable set of plays, are uniquely defined~\cite{Var85}.
For a state $q \in Q$ and an event $\Aa\subseteq\Paths$, we denote by
$\Prb_q^{\straa}(\Aa)$ the probability that a play belongs 
to $\Aa$ if the game starts from the state $q$ and player~$1$ follows
the strategy~$\straa$.
For a measurable function $f:\Paths \to \real$ we denote by 
$\Exp_q^{\straa}[f]$ the \emph{expectation} of the function
$f$ under the probability measure $\Prb_q^{\straa}(\cdot)$.

Strategies that do not use randomization are called pure.
A player-1 strategy~$\straa$ is \emph{pure} if for all $\rho \in Q^*$
and $q \in Q_1$, there is a state~$q' \in Q$ such that  
$\straa(\rho \cdot q)(q') = 1$.

\smallskip\noindent{\bf Finite-memory strategies.}
A strategy uses \emph{finite-memory} if it can be encoded
by a deterministic transducer $\tuple{\Mem, m_0, \alpha_u, \alpha_n}$
where $\Mem$ is a finite set (the memory of the strategy), 
$m_0 \in \Mem$ is the initial memory value,
$\alpha_u: \Mem \times Q \to \Mem$ is an update function, and 
$\alpha_n: \Mem \times Q_1 \to \dist(Q)$ is a next-move function. 
The \emph{size} of the strategy is the number $\abs{\Mem}$ of memory values.
If the game is in a player-$1$ state $q$, and $m$ is the current memory value,
then the strategy chooses the next state $q'$ according to 
the probability distribution $\alpha_n(m,q)$,
and the memory is updated to $\alpha_u(m,q)$. 
Formally, $\tuple{\Mem, m_0, \alpha_u, \alpha_n}$
defines the strategy $\straa$ such that $\straa(\rho\cdot q) = \alpha_n(\hat{\alpha}_u(m_0, \rho), q)$
for all $\rho \in Q^*$ and $q \in Q_1$, where $\hat{\alpha}_u$ extends $\alpha_u$ to sequences
of states as expected. A strategy is \emph{memoryless} if $\abs{\Mem} = 1$.
For a finite-memory strategy $\straa$, let $M_{\straa}$ be the Markov chain 
obtained as the product of $M$ with the transducer defining $\straa$, 
where $(\tuple{m,q},\tuple{m',q'})$ is an edge
in $M_{\straa}$ if $m' = \alpha_u(m,q)$ and either $q \in Q_1$ and 
$q' \in \Supp(\alpha_n(m,q))$, or $q \in Q_P$ and $(q,q') \in E$.

\paragraph{\bf Two-player games.}
A \emph{two-player game} is a graph $G =(Q, E)$ with the same assumptions
as for MDP, except that the partition of $Q$ is denoted $(Q_1,Q_2)$ where
$Q_2$ is the set of \mbox{\emph{player-$2$}} \emph{states}. The notions of 
play, strategies (in particular strategies for player~$2$), and outcome are 
analogous to the case of MDP~\cite{CD10a}.

\smallskip\noindent{\bf Objectives.}
An \emph{objective} for an MDP $M$ (or game~$G$) is a set $\phi \subseteq \Paths$
of infinite paths. 
Let $p:Q \to \nat$ be a \emph{priority function} and $w:E \to \zed$ be a 
\emph{weight function}\footnote{Sometimes we take the freedom to use rational weights (i.e., $\weight:E \to \rat$), 
while we always assume that weights are integers encoded in binary for complexity results.}  
where positive numbers represent rewards. 
We denote by $W$ the largest weight (in absolute value) according to~$\weight$.
The \emph{energy level} of a prefix $\gamma = q_0 q_1 \dots q_n$ of a play
is $\EL(\weight,\gamma) = \sum_{i=0}^{n-1} \weight(q_i,q_{i+1})$, and the \emph{mean-payoff value}\footnote{The results 
of this paper hold for the definition of mean-payoff value using $\limsup$ instead of $\liminf$.}
of a play $\rho= q_0 q_1 \dots$ is $\MP(\weight,\rho) = \liminf_{n \to \infty} \frac{1}{n}\cdot\EL(\weight,\rho(n))$. 
In the sequel, when the weight function~$\weight$ is clear from the context we omit it and
simply write $\EL(\gamma)$ and $\MP(\rho)$. 
We denote by $\Inf(\rho)$ the set of states that occur infinitely often in $\rho$,
and we consider the following objectives:

\begin{compactitem}
 	\item \emph{Parity objectives.}
	The \emph{parity} objective $\Parity(p) = \{\rho \in \Paths \mid \min\{p(q) \mid q \in \Inf(\rho)\} \text{ is even }\}$
	requires that the minimum priority visited infinitely often be even. 
	The special cases of \emph{B\"uchi} and \emph{coB\"uchi} objectives correspond
	to the case with two priorities, $p: Q \to \{0,1\}$ and $p: Q \to \{1,2\}$ respectively.

	\item \emph{Energy objectives.}
	Given an initial credit $c_0 \in \nat$, the \emph{energy} objective 
	$\PosEnergy(c_0) = \{ \rho \in \Paths \mid \forall n \geq 0:  c_0 + \EL(\rho(n)) \geq 0 \}$
	requires that the energy level be always positive.

	\item \emph{Mean-payoff objectives.}
	Given a threshold $\nu \in \rat$, the \emph{mean-payoff} objective 
	$\MeanPayoff^{\geq \nu} = \{ \rho \in \Paths \mid \MP(\rho) \geq \nu \}$
	(resp. $\MeanPayoff^{> \nu} = \{ \rho \in \Paths \mid \MP(\rho) > \nu \}$)
	requires that the mean-payoff value be at least $\nu$ (resp. strictly greater than $\nu$).

	\item \emph{Combined objectives.}
	The \emph{energy parity} objective $\Parity(p) \cap \PosEnergy(c_0)$
	and the \emph{mean-payoff parity} objective $\Parity(p) \cap \MeanPayoff^{\sim \nu}$ 
        (for $\sim \in \{\geq,>\}$)
	combine the requirements of parity and energy (resp., mean-payoff) objectives.
\end{compactitem}


\smallskip\noindent{\bf Almost-sure winning strategies.}
For MDPs, we say that a player-$1$ strategy~$\straa$ is 
\emph{almost-sure winning} 
in a state~$q$ for an objective~$\phi$ if 
$\Prb_q^{\straa}(\phi) = 1$.
For two-player games, we say that a player-$1$ strategy~$\straa$ is \emph{winning}
in a state~$q$ for an objective~$\phi$ if all outcomes of $\straa$ starting in~$q$
belong to $\phi$.
For energy objectives with unspecified initial credit, 
we also say that a strategy is (almost-sure) winning if 
it is (almost-sure) winning for \emph{some} finite initial credit. 



\smallskip\noindent{\bf Decision problems.}
We are interested in the following problems. Given an 
MDP $M$ with weight function $\weight$ and priority function $p$, 
and a state $q_0$,
\begin{compactitem}
\item the \emph{energy parity problem} asks whether there exists a finite 
initial credit $c_0 \in \nat$ and an 
almost-sure winning strategy for the energy parity objective from~$q_0$ with initial credit~$c_0$.
We are also interested in computing the \emph{minimum initial credit} in~$q_0$
which is the least value of initial credit for which there exists an almost-sure winning strategy 
for player~$1$ in $q_0$.
A strategy for player~$1$ is \emph{optimal} in $q_0$ if it is winning 
from $q_0$ with the minimum initial credit;

\item the \emph{mean-payoff parity problem} asks whether there exists an 
almost-sure winning strategy for the mean-payoff parity objective with threshold~$0$
from~$q_0$.
Note that it is not restrictive to consider mean-payoff objectives with threshold~$0$
because for $\sim \in \{\geq,>\}$, we have $\MP(\weight,\rho) \sim \nu$ 
iff $\MP(\weight-\nu,\rho) \sim 0$, 
where $\weight-\nu$ is the weight function that assigns $\weight(e) - \nu$ to each 
edge $e \in E$.

\end{compactitem}

The two-player game versions of these problems are defined analogously~\cite{CD10a}.
It is known that the initial credit problem for simple two-player energy games~\cite{CAHS03,BFLMS08},
as well as for two-player parity games~\cite{EJ91}
can be solved in NP~$\cap$~coNP because memoryless strategies are sufficient to win.
Moreover, parity games reduce in polynomial time to mean-payoff games~\cite{Jurdzinski98}, 
which are log-space equivalent to energy games~\cite{BFLMS08,BCDGR10}. It is a long-standing open 
question to know if a polynomial-time algorithm exists for these problems.
Finally, energy parity games and mean-payoff parity games are solvable in NP~$\cap$~coNP
although winning strategies may require exponential and infinite memory respectively,
even in one-player games (and thus also in MDPs)~\cite{ChatterjeeHJ05,CD10a}.

The decision problem for MDPs with parity objective, as well as with mean-payoff objective, 
can be solved in polynomial time~\cite{FV97,CY95,CH11,deAlfaro97}. However, the problem is in NP~$\cap$~coNP 
for MDPs with energy objective because an MDP with
energy objective is equivalent to a two-player energy game (where the probabilistic
states are controlled by player~$2$). Indeed $(1)$ a winning strategy in the game
is trivially almost-sure winning in the MDP, and $(2)$ if an almost-sure winning 
strategy $\straa$ in the MDP was not winning in the game, then for all initial
credit $c_0$ there would exist an outcome $\rho$ of $\straa$ such that $c_0 + \EL(\rho(i)) < 0$
for some position $i \geq 0$. The prefix $\rho(i)$ has a positive probability 
in the MDP, in contradiction with the fact that $\straa$ is almost-sure winning.
As a consequence, solving MDP with energy objectives is at least as hard as 
solving parity games.

In this paper, we show that the decision problem for MDPs with energy parity objective
is in NP~$\cap$~coNP, which is the best conceivable upper bound unless parity games
can be solved in P. And for MDPs with mean-payoff parity objective, we show that the 
decision problem can be solved in polynomial time, improving a recent PSPACE 
bound~\cite{GIMBERT:2011:HAL-00559173:2}.

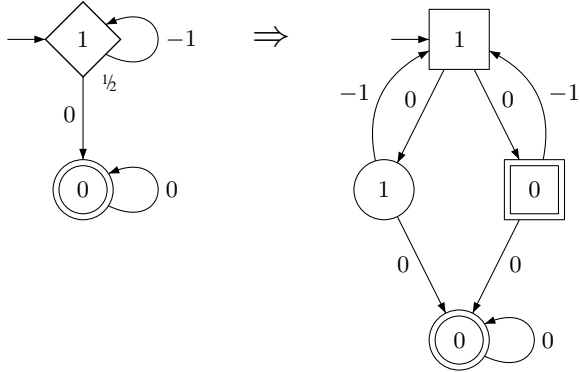
\begin{figure}[!tb]
  \begin{center}
    \hrule
    \begin{picture}(82,55)(0,0)



\rpnode[Nmarks=i](p1)(10,47)(4,5){$1$}
\rpnode[Nmarks=n, ExtNL=y, NLangle=300, NLdist=1.5](p1)(10,47)(4,5){$\sfrac{1}{2}$}
\node[Nmarks=r, Nframe=y, rdist=0.8](p2)(10,27){$0$}   

\drawloop[ELside=r,loopCW=n, loopangle=0, loopheight=5, loopwidth=6](p1){$-1$}
\drawedge[ELpos=50, ELside=r, curvedepth=0](p1,p2){$0$}
\drawloop[ELside=r,loopCW=n, loopangle=0, loopdiam=6](p2){$0$}

\node[Nframe=n](label)(35,47){{\Large $\Rightarrow$}}

\node[Nmarks=i, Nmr=0](q1)(60,47){$1$}
\node[Nmarks=n](q2)(50,27){$1$}
\node[Nmarks=r, Nmr=0, rdist=0.8](q3)(70,27){$0$}
\node[Nmarks=r, Nframe=y, rdist=0.8](q4)(60,7){$0$}


\drawedge[ELpos=45, ELside=r, curvedepth=0](q1,q2){$0$}
\drawedge[ELpos=45, ELside=l, curvedepth=0](q1,q3){$0$}

\drawedge[ELpos=45, ELside=l, curvedepth=6](q2,q1){$-1$}
\drawedge[ELpos=45, ELside=r, curvedepth=-6](q3,q1){$-1$}

\drawedge[ELpos=45, ELside=r, curvedepth=0](q2,q4){$0$}
\drawedge[ELpos=45, ELside=l, curvedepth=0](q3,q4){$0$}

\drawloop[ELside=r,loopCW=n, loopangle=0, loopdiam=6](q4){$0$}


\end{picture}
    \hrule
    \caption{The gadget construction is wrong for mean-payoff parity MDPs. 
Player~$1$ is almost-sure winning for mean-payoff B\"uchi in the MDP (on the left)
but player~$1$ is losing in the two-player game (on the right)
because player~$2$ (box-player) can force a negative-energy cycle.  \label{fig:energy-vs-mean-payoff}}
  \end{center}
\vspace{-2em}
\end{figure}

The MDP in \figurename~\ref{fig:energy-vs-mean-payoff} on the left, which is essentially a Markov chain,
is an example where the mean-payoff parity condition is satisfied almost-surely,
while the energy parity condition is not, no matter the value of the initial credit. 
For initial credit $c_0$, the energy will drop below $0$ with positive probability, namely $\frac{1}{2^{c_0+1}}$.





\smallskip\noindent{\bf End-component lemma.}
We now present an important lemma about end-components from~\cite{CY95,deAlfaro97}
that we use in the proofs of our result.
It states that for arbitrary strategies (memoryless or not), 
with probability 1 the set of states visited infinitely often along a play is 
an end-component.
This lemma allows us to derive conclusions on the (infinite) set of
plays in an MDP by analyzing the (finite) set of end-components in the MDP.

\begin{lemma}\label{lemm:end-component}
  \cite{CY95,deAlfaro97} Given an MDP $M$, for all states $q \in Q$
  and all strategies $\straa \in \Straa$, we have
  $\Prb_q^{\straa}(\set{\pat \mid \Inf(\pat) \in \cale(M)})=1$.
\end{lemma}

\section{MDPs with Energy Parity Objectives}

We show that energy parity MDPs can be solved in NP~$\cap$~coNP, using
a reduction to two-player energy B\"uchi games. 
Our reduction also preserves the value of the minimum initial credit.
Therefore, we obtain a pseudo-polynomial algorithm for this problem,
which also computes the minimum initial credit. 
Moreover, we show that the memory requirement for almost-sure winning strategies 
is at most $2 \! \cdot \! \abs{Q} \! \cdot \! W$, which is essentially 
optimal\footnote{Example~$1$ in~\cite{CD10a} shows that memory of size 
$2 \! \cdot \! (\abs{Q}-1) \! \cdot \! W + 1$ may be necessary.}.

We first establish the results for the special case of energy B\"uchi MDPs. 
We present a reduction of the energy B\"uchi problem for MDPs 
to the energy B\"uchi problem for two-player games. The result
then follows from the fact that the latter problem is in NP~$\cap$~coNP
and solvable in pseudo-polynomial time~\cite{CD10a}.

\begin{figure}[!tb]
  \begin{center}
    \hrule
    \begin{picture}(85,55)(0,0)



\rpnode[Nmarks=n](p1)(10,45)(4,5){$1$}
\rpnode[Nmarks=n, ExtNL=y, NLangle=135, NLdist=1](p1)(10,45)(4,5){$q$}
\rpnode[Nmarks=n, ExtNL=y, NLangle=270, NLdist=1.5](p1)(10,45)(4,5){$\sfrac{1}{2}$}
\node[Nmarks=n, Nframe=n, Nw=5,Nh=5](dummy1)(2,30){$\cdot$}
\node[Nmarks=n, Nframe=n, Nw=5,Nh=5](dummy2)(18,30){$\cdot$}

\drawedge[ELpos=50, ELside=r, curvedepth=0](p1,dummy1){$\gamma_1$}
\drawedge[ELpos=50, ELside=l, curvedepth=0](p1,dummy2){$\gamma_2$}

\node[Nframe=n](label)(32,45){{\Large $\Rightarrow$}}

\node[Nmarks=n, Nmr=0](q1)(60,45){$1$}
\node[Nmarks=n, Nmr=0, ExtNL=y, NLangle=145, NLdist=1](q1)(60,45){$q$}
\node[Nmarks=n](q2)(50,25){$1$}
\node[Nmarks=n, ExtNL=y, NLangle=140, NLdist=1](q2)(50,25){$(q,\L)$}
\node[Nmarks=r, Nmr=0, rdist=0.8](q3)(70,25){$0$}          
\node[Nmarks=n, Nmr=0, ExtNL=y, NLangle=35, NLdist=1.4](q3)(70,25){$(q,\R)$}

\node[Nmarks=n, Nframe=n, Nw=5,Nh=5](dummy1)(50,5){$\cdot$}
\node[Nmarks=n, Nframe=n, Nw=5,Nh=5](dummy2)(70,5){$\cdot$}


\drawedge[ELpos=50, ELside=r, curvedepth=0](q1,q2){$0$}
\drawedge[ELpos=50, ELside=l, curvedepth=0](q1,q3){$0$}

\drawedge[ELpos=40, ELside=r, curvedepth=0](q2,dummy1){$\gamma_1$}
\drawedge[ELpos=27, ELside=l, ELdist=0, curvedepth=0](q2,dummy2){$\gamma_2$}

\drawedge[ELpos=27, ELside=r, ELdist=0, curvedepth=0](q3,dummy1){$\gamma_1$}
\drawedge[ELpos=40, ELside=l, curvedepth=0](q3,dummy2){$\gamma_2$}


\end{picture}
    \hrule
    \caption{Gadget for probabilistic states in energy B\"uchi MDP. 
Diamonds are probabilistic states, circles are player~$1$ states,
and boxes are player~$2$ states. \label{fig:gadget-Buchi}}
  \end{center}
\vspace{-2em}
\end{figure}
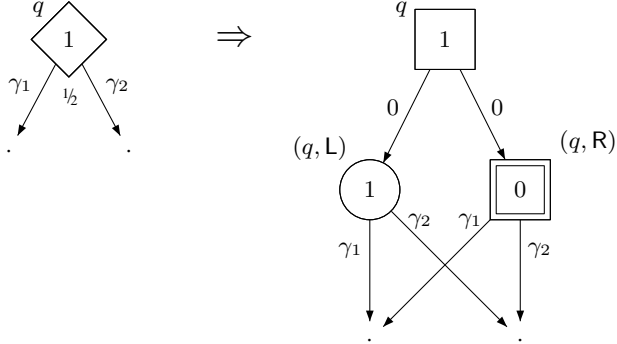

Given an MDP $M$, we can assume without loss of generality that  
every probabilistic state has priority~$1$, and has two outgoing transitions
with probability$\,\sfrac{1}{2}$ each~\cite[Section~6]{ZwickP96}. 
We construct a two-player game $G$ by replacing every probabilistic state of $M$
by a gadget as in \figurename~\ref{fig:gadget-Buchi}. The probabilistic states $q$ 
of $M$ are mapped to player-$2$ states in $G$ with two successors $(q,\L)$ and $(q,\R)$. 
Intuitively, player~$2$ chooses $(q,\L)$ to check whether player~$1$ can enforce 
the B\"uchi condition almost-surely. This is the case if player~$1$ can reach 
a B\"uchi state (with priority~$0$) infinitely often when he controls the 
probabilistic states (otherwise, no B\"uchi state is ever visited, and 
since $(\cdot,\L)$ states have priority~$1$, the B\"uchi condition is not realized in~$G$).
And player~$2$ chooses $(q,\R)$ to check that the energy condition is satisfied.
If player~$2$ can exhaust the energy level in $G$, then the corresponding play
prefix has positive probability in~$M$. Note that $(q,\R)$ has priority~$0$
and thus cannot be used by player~$2$ to spoil the B\"uchi condition.

Formally, given $M = (Q, E, \trans)$ with partition $(Q_1,Q_P)$ of $Q$,
we construct a game $G = (Q',E')$ with partition $(Q'_1,Q'_P)$ where
$Q'_1 =  Q_1 \cup (Q_P \times \{\L\})$ and 
$Q'_2 =  Q_P \cup (Q_P \times \{\R\})$, see also \figurename~\ref{fig:gadget-Buchi}. 
The states in $Q'$ that are already in $Q$ get the same priority as in $M$,
the states $(\cdot,\L)$ have priority $1$, and the states $(\cdot,\R)$ have priority $0$. 
The set $E'$ contains the following edges:
\begin{compactitem}
\item all edges $(q,q') \in E$ such that $q \in Q_1$; 
\item edges $(q,(q,d))$, $((q,d),q')$ for all $q \in Q_P$, $d \in \{\L,\R\}$, and $q' \in \Supp(\trans(q))$.
\end{compactitem}
\smallskip
The edges $(q,q')$ and $((q,d),q')$ in $E'$ get the same weight as $(q,q')$ in $M$,
and all edges $(q,(q,d))$ get weight $0$.

\begin{lemma}\label{lem:energy-buchi-MDP}
Given an MDP $M$ with energy B\"uchi objective, we can construct in linear
time a two-player game $G$ with energy B\"uchi objective such that 
for all states $q_0$ in $M$, there exists an almost-sure winning strategy from
$q_0$ in $M$ if and only if there exists a winning strategy from~$q_0$ in $G$
(with the same initial credit).
\end{lemma}

Note that the reduction presented in the proof of Lemma~\ref{lem:energy-buchi-MDP}
would not work for mean-payoff B\"uchi MDPs. Consider the MDP on 
\figurename~\ref{fig:energy-vs-mean-payoff} for which the gadget-based reduction
to two-player games is shown on the right. The game is losing for player~$1$ 
both for energy and mean-payoff parity, simply because player~$2$ can always choose
to loop through the box states, thus realizing a negative energy and mean-payoff
value (no matter the initial credit). However player~$1$ is almost-sure winning
in the mean-payoff parity MDP (on the left in \figurename~\ref{fig:energy-vs-mean-payoff}).

While the reduction in the proof of Lemma~\ref{lem:energy-buchi-MDP} 
gives a game with $n' = \abs{Q_1}+3\cdot\abs{Q_P}$ states, the structure
of the gadgets (see \figurename~\ref{fig:gadget-Buchi}) is such that the
energy level is independent of which of the transitions $(q,(q,\L))$ or
$(q,(q,\R))$ is taken. Since from the result of~\cite[Lemma~8]{CD10a} and its proof, 
it follows that the memory updates in winning strategies for energy B\"uchi games 
can be done according to the energy level of the play prefix, it follows that
the memory bound of $2 \! \cdot \! n \! \cdot \! W$ can be transfered
to almost-sure winning strategies in Energy B\"uchi MDPs, where 
$n = \abs{\Win \cap Q_1}$ is the number of player~$1$ almost-sure winning states.
Also, the pseudo-polynomial algorithm for solving two-player energy 
B\"uchi games can be used for MDPs, with the same $O(\abs{E} \cdot \abs{Q}^{5} \cdot W)$ complexity~\cite[Table~1]{CD10a} .

\medskip
Using Lemma~\ref{lem:energy-buchi-MDP}, we solve energy parity MDPs by 
a reduction to energy B\"uchi MDPs. 
The key idea of the reduction is that if player~$1$ has an almost-sure 
winning strategy for the energy parity objective, then player~1 can choose 
an even priority~$2i$ and decide to satisfy the energy objective 
along with satisfying that priority~$2i$ is visited infinitely 
often, and priorities less than~$2i$ are visited finitely often.

W.l.o.g. we assume that player-$1$ states
and probabilistic states alternate, i.e. $E(q) \subseteq Q_1$ for
all $q \in Q_P$, and $E(q) \subseteq Q_P$ for all $q \in Q_1$.
The reduction is then as follows. Given an MDP $M =(Q, E, \trans)$
with a priority function $p:Q \to \nat$ and a weight function $\weight:E \to \zed$,
we construct $\tuple{M',p',\weight'}$ as follows. 
$M'$ is the MDP $M =(Q', E', \trans')$ where:
\begin{compactitem}
\item $Q' = Q \cup (Q \times \{0,2,\dots,2r\}) \cup \{\sink\}$ where $2r$ is the largest even
priority of a state in $Q$. Intuitively, a state $(q,i) \in Q'$ corresponds 
to the state $q$ of $M$ from which player~$1$ will ensure to visit priority $i$ (which is even)
infinitely often, and never visit priority smaller than~$i$;
\item $E'$ contains $E \cup \{(\sink,\sink)\}$ and the following edges. 
For each probabilistic state $q \in Q_P$, for $i=0,2,\dots,2r$, 
\begin{compactitem}
\item $(a)$ if $p(q') \geq i$ for all $q' \in E(q)$, then $((q,i),(q',i)) \in E'$ for all $q' \in E(q)$, 
\item $(b)$ otherwise, $((q,i),\sink) \in E'$.
\end{compactitem}
For each player~$1$ state $q \in Q_1$, for each $q' \in E(q)$, for $i=0,2,\dots,2r$, 
\begin{compactitem}
\item $(a)$ $(q, \sink) \in E'$ and $((q,i),\sink) \in E'$, and
\item $(b)$ if $p(q') \geq i$, then $(q,(q',i)) \in E'$ and $((q,i),(q',i)) \in E'$.
\end{compactitem}
\end{compactitem}
\smallskip
The partition $(Q'_1,Q'_P)$ of $Q'$ is defined by $Q'_1 = Q_1 \cup (Q_1 \times \{0,2,\dots,2r\}) \cup \{\sink\}$
and $Q'_P = Q' \setminus Q'_1$.
The weight of the edges $(q,q')$, $(q,(q',i))$ and $((q,i),(q',i))$ according to $\weight'$ 
is the same as the weight of $(q,q')$ according to $\weight$. 
The states $(q,i)$ such that $p(q) = i$ have priority $0$ according to $p'$ 
(they are the B\"uchi states), and all the other states in $Q'$ 
(including $\sink$) have priority $1$.

\begin{lemma}\label{lem:energy-parity-MDP}
Given an MDP $M$ with energy parity objective, we can construct in quadratic 
time an MDP $M'$ with energy B\"uchi objective such that 
for all states $q_0$ in $M$, 
there exists an almost-sure winning strategy from $q_0$ in $M$ 
if and only if there exists an almost-sure winning strategy from~$q_0$ in $M'$ 
(with the same initial credit).
\end{lemma}

From the proof of Lemma~\ref{lem:energy-parity-MDP}, it follows that the 
memory requirement is the same as for energy B\"uchi MDPs. 
And if the weights are in $\{-1,0,1\}$, it follows that the energy parity problem
can be solved in polynomial time.

\begin{theorem}\label{theo:energy-parity-MDP}
For energy parity MDPs, 
\begin{compactenum}
\item the decision problem of whether a given state is almost-sure winning is in NP~$\cap$~coNP,
and there is a pseudo-polynomial time algorithm in $O(\abs{E} \cdot d \cdot \abs{Q}^{5} \cdot W)$
to solve it;
\item memory of size $2 \! \cdot \! \abs{Q} \! \cdot \! W$ 
 is sufficient for almost-sure winning strategies.
\end{compactenum}
\end{theorem}

\section{MDPs with Mean-payoff Parity Objectives}\label{sec:MDP-mean-payoff-parity}
In this section we present a polynomial-time algorithm for 
solving MDPs with mean-payoff parity objective.
We first recall some useful properties of MDPs.
 
For an end-component $U \in \cale(M)$, consider the memoryless
strategy $\straa_U$ that plays in every state $s \in U \cap Q_1$ all
edges in $E(s) \cap U$ uniformly at random.  Given the strategy
$\straa_U$, the end-component $U$ is a closed connected recurrent set
in the Markov chain obtained by fixing $\straa_U$.

\begin{lemma}\label{lemm:end-component1}
  Given an MDP $M$ and an end-component $U \in \cale(M)$, the strategy
  $\straa_U$ ensures that for all states $s \in U$, we have
  $\Prb_s^{\straa_U}(\set{\pat \mid \Inf(\pat)=U})=1$.
\end{lemma}

\noindent{\bf Expected mean-payoff value.} Given an MDP $M$ 
with a weight function $\weight$, the \emph{expected mean-payoff value}, 
denoted $\vaMP(\weight)$, is the function that assigns to every 
state the maximal expectation of the mean-payoff objective 
that can be guaranteed by any strategy.
Formally, for $q \in Q$ we have 
$\vaMP(\weight)(q)= 
\sup_{\straa \in \Straa} \Exp_{q}^{\straa}(\MP(w))$, where $\MP(w)$ is the 
measurable function that assigns to a play $\rho$ the long-run average 
$\MP(w,\rho)$ of the weights.
By the classical results of MDPs with mean-payoff objectives, 
it follows that there exists pure memoryless optimal strategies~\cite{FV97},
i.e., there exists a pure memoryless optimal strategy $\straa^*$ such 
that for all $q \in Q$ we have $\vaMP(\weight)(q) = \Exp_{q}^{\straa^*}(\MP(w))$.

It follows from Lemma~\ref{lemm:end-component1} 
that the strategy $\straa_U$ ensures that from any
starting state~$s$, any other state~$t$ is reached in finite time
with probability~1. Therefore,
the value for mean-payoff parity objectives in MDPs
can be obtained by computing values for end-components and then
playing a strategy to maximize the expectation to reach the 
values of the end-components.

We now present the key lemma where we show that for an MDP that is an 
end-component such that the minimum priority is even, the mean-payoff 
parity objective $\Parity(p) \cap \MeanPayoff^{\geq \nu}$ is satisfied 
with probability~1 if the expected mean-payoff value is at least $\nu$ at 
some state (the result also holds for strict inequality).
In other words, from the expected mean-payoff value of at least $\nu$ 
we ensure that both the mean-payoff and parity objective is satisfied with 
probability~1 from all states.
The proof of the lemma considers two pure memoryless strategies: one 
for stochastic shortest path and the other for optimal expected mean-payoff 
value, and combines them to obtain an almost-sure winning strategy for 
the mean-payoff parity objective (details in appendix).

\begin{lemma}\label{lemm:key}
  Consider an MDP $M$ with state space $Q$, a priority function $p$,
  and weight function $\weight$ such that (a)~$M$ is an end-component (i.e.,
  $Q$ is an end-component) and (b)~the smallest priority in~$Q$ is
  even. 
  If there is a state $q \in Q$ such that 
  $\vaMP(\weight)\geq\nu$ (resp. $\vaMP(\weight)>\nu$),  
  then there exists a strategy $\straa^*$ such that for all states $q \in  Q$
  we have $\Prb_q^{\straa^*}(\Parity(p) \cap \MeanPayoff^{\geq \nu})=1$ 
  (resp. $\Prb_q^{\straa^*}(\Parity(p) \cap \MeanPayoff^{> \nu})=1$). 
\end{lemma}

\noindent{\bf Memory required by strategies.}
Lemma~\ref{lemm:key} shows that if the smallest priority in an end-component
is even, then considering the sub-game restricted to the end-component,
the mean-payoff parity objective is satisfied
if and only if the mean-payoff objective is satisfied.
The strategy constructed in Lemma~\ref{lemm:key} requires infinite memory,
and in the case of loose inequality (i.e., $\MeanPayoff^{\geq \nu}$) infinite memory is required
in general (see~\cite{ChatterjeeHJ05} for an example on graphs),
and if the inequality is strict (i.e., $\MeanPayoff^{> \nu}$), then finite memory 
strategies exist~\cite{GIMBERT:2011:HAL-00559173:2}.
For the purpose of computation we show that both strict and 
non-strict inequality can be solved in polynomial time. 
Since Lemma~\ref{lemm:key} holds for both strict and non-strict 
inequality, in sequel of this section we consider non-strict inequality 
and all the results hold for non-strict inequality as well.

\smallskip\noindent{\bf Winning end-component.} 
Given an MDP $M$ with a parity objective $\Parity(p)$ and a
mean-payoff objective $\MeanPayoff^{\geq \nu}$ for a weight function 
$\weight$, we call an 
end-component $U$ \emph{winning} if  
(a)~$\min(p(U))$ is even; and 
(b)~there exists a state with expected mean-payoff value at least $\nu$ 
in the sub-MDP induced by $U$, i.e., $\max_{q \in U} 
\vaMP(\weight)(q) \geq \nu$ in the sub-MDP induced by $U$.
We denote by $\calw$ the set of winning end-components, and 
let $\Win=\bigcup_{U \in \calw} U$ be the union of the winning end-components.

\smallskip\noindent{\bf Reduction to reachability of winning 
end-component.} 
By Lemma~\ref{lemm:key} it follows that in every winning end-component
the mean-payoff parity objective is satisfied with probability~1.
Conversely, consider an end-component $U$ that is not winning, then 
either the smallest priority is odd, or the maximal expected mean-payoff 
value that can be ensured for any state in $U$ by staying in $U$ 
is less than $\nu$. 
Hence if only states in $U$ are visited infinitely often, then
with probability~1 (i)~either the parity objective is not satisfied,
or (ii)~the mean-payoff objective is not satisfied.
In other words, if an end-component that is not winning is 
visited infinitely often, then the mean-payoff parity objective 
is satisfied with probability~0. 
It follows that the value function for MDPs with mean-payoff parity
objective can be computed by computing the value function for reachability 
to the set $\Win$, i.e., formally,
$\sup_{\straa\in\Straa}\Prb_{q}^{\straa}(\Parity(p) \cap \MeanPayoff^{\geq \nu})
=\sup_{\straa \in \Straa} \Prb_{q}^{\straa}(\Reach(\Win))$, 
where $\Reach(\Win)$ is the set of paths that reaches a state in $\Win$ at least
once.
Since the value function in MDPs with reachability objectives can 
be computed in polynomial time using linear programming~\cite{FV97}, 
it suffices to present a polynomial-time algorithm to compute~$\Win$
in order to obtain a polynomial-time algorithm for MDPs with 
mean-payoff parity objectives.

\smallskip\noindent{\bf Computing winning end-components.}  
The computation of the winning end-components is done iteratively 
by computing winning end-components with smallest priority~0, 
then winning end-components with smallest priority~2, and so on.
The computation of $\Win$ is as follows:
\begin{compactitem}
\item For $i\geq 0$, let $\calw_{2i}$ be the set of maximal end-components $U$ 
  with states with priority at least $2i$ and that contain
  at least one state with priority~$2i$, i.e., $U$ contains only states with
  priority at least $2i$, and contains at least one state with priority $2i$.
  Let $\calw'_{2i} \subseteq \calw_{2i}$ be the set of maximal end-components 
  $U \in \calw_{2i}$ such that there is a state $q \in U$ such that the 
  expected mean-payoff value in the sub-MDP restricted to $U$ is at least 
  $\nu$. Let $\Win_{2i}= \bigcup_{U \in \calw'_{2i}} U$.
\end{compactitem}
The set $\Win = \bigcup_{i=0}^{\lfloor d/2 \rfloor} \Win_{2i}$ is the union 
of the states of the winning end-components (formal pseudo-code in the 
appendix). 

\smallskip\noindent{\bf Complexity of computing winning end-components.} 
The winning end-component algorithm runs for $O(d)$ iterations and in each 
iteration requires to compute a maximal end-component decomposition and 
compute mean-payoff values of at most $n$ end-components, where $n$ is 
the number of states of the MDP. 
The maximal end-component decomposition can be achieved in polynomial 
time~\cite{CY95,deAlfaro97,CH11}.
The mean-payoff value function of an MDP can also be computed in 
polynomial time using linear programming~\cite{FV97}.
It follows that the value function of an MDP with mean-payoff parity 
objectives can be computed in polynomial time.
The almost-sure winning set is obtained by computing almost-sure 
reachability to $\Win$ in polynomial time~\cite{CY95,deAlfaro97,CH11}.
This polynomial-time complexity provides a tight upper bound for the problem, 
and closes the gap left by the PSPACE upper bound of~\cite{GIMBERT:2011:HAL-00559173:2}.

\begin{theorem}
The following assertions hold:
\begin{compactenum}
\item The set of almost-sure winning states for mean-payoff parity 
objectives can be computed in polynomial time for MDPs.
\item For mean-payoff parity objectives, almost-sure winning strategies
require infinite memory in general for non-strict inequality (i.e, for 
mean-payoff parity objectives $\Parity(p) \cap \MeanPayoff^{\geq \nu}$) and 
finite-memory almost-sure winning strategies exist for strict inequality 
(i.e., for $\Parity(p) \cap \MeanPayoff^{>\nu}$). 
\end{compactenum}
\end{theorem}

\noindent{\bf Discussion.} 
We considered MDPs with conjunction of mean-payoff parity and 
energy parity objectives, and presented tight complexity bounds, algorithms,
and bounds for the memory required by strategies. 
The disjunction of mean-payoff parity and energy parity 
objectives are straightforward and summarized in Theorem~\ref{thrm:disjunction} 
(details in appendix).

\begin{theorem}\label{thrm:disjunction}
The following assertions hold:
\begin{compactenum}
\item The set of almost-sure winning states for disjunction of mean-payoff
and parity objectives can be computed in polynomial time for MDPs.
\item The decision problem of whether a given state is almost-sure winning
for disjunction of energy and parity objectives is in NP $\cap$ coNP for MDPs.
\end{compactenum}
\end{theorem}




\newpage
\section*{Appendix}


\section{Details of Section~3}

\begin{proof}[of Lemma~\ref{lem:energy-buchi-MDP}]
We show that 
player~$1$ has an almost-sure winning strategy in $M$ if and only if 
player~$1$ has a winning strategy in the game $G$ (for the same initial credit). 

First, we show that if player~$1$ has an almost-sure winning strategy $\straa_M$ 
in $M$, then we can construct a winning strategy $\straa_G$ in $G$.
We can assume that $\straa_M$ is pure~\cite[Theorem~5]{CDGH10}.

To define $\straa_G$, we assign a rank to prefixes of outcomes of 
$\straa_M$ in $M$ as follows. Prefixes $\rho$ such that $p(\Last(\rho)) = 0$
get rank $0$. For other prefixes~$\rho$ (with $p(\Last(\rho)) = 1$), if $\Last(\rho) \in Q_1$ is a
player-$1$ state, then $\rho$ gets rank $1 + \rank(\rho \cdot q)$
where $q$ is such that $\straa(\rho)(q) = 1$; 
if $\Last(\rho) \in Q_P$ is a probabilistic state, then $\rho$ gets rank 
$1 + \min \{ \rank(\rho') \mid \rho' \text{ is a ranked successor of } \rho\}$.
Prefixes without ranked successor get no rank.
We claim that all prefixes $\rho$ compatible with $\straa_M$ get a (finite) rank.
Otherwise, there would exist a non-ranked prefix compatible with $\straa_M$
(thus reachable with positive probability) such that all its extensions 
are unranked. This would imply that only states with priority~$1$ are visited
from that point on, hence the co-B\"uchi objective has positive probability,
in contradiction with the fact that $\straa_M$ is almost-sure winning
for energy B\"uchi.

We construct the pure strategy $\straa_G$ as follows.
Given a play prefix $\rho_G$ in $G$, let $h(\rho_G)$ be the sequence
obtained from $\rho_G$ by deleting all states of the form $(q,d)$
for $q \in Q_P$ and $d \in \{\L,\R\}$. Note that $h(\rho_G)$ is a play in $M$.
Let $q_G = \Last(\rho_G) \in Q'_1$, we define $\straa_G(\rho_G)$ as follows:
\begin{itemize}
\item if $q_G \in Q_1$, then $\straa_G(\rho_G) = \straa_M(h(\rho_G))$;
\item if $q_G = (q,\L)$ (for $q \in Q_P$), 
then $\straa_G(\rho_G) = q'$ where $\rank(\rho_G \cdot q') < \rank(\rho_G)$.
\end{itemize}
Note that for every outcome $\rho_G$ of $\straa_G$, the play $h(\rho_G)$
is an outcome of $\straa_M$ in~$M$.
Towards contradiction, assume that $\straa_G$ is not winning in $G$.
Then, there exists an outcome $\rho_G$ of $\straa_G$ that violates either:
\begin{itemize}
\item the energy condition; then, the energy level drops below $0$
after finitely many steps in $\rho_G$, and this occurs as well in 
$h(\rho_G)$ with positive probability in $M$, a contradiction with
the fact that $\straa_M$ is almost-sure winning for energy B\"uchi in $M$.

\item or the B\"uchi condition; then, from some  point on in $\rho_G$
only priority~$1$ is visited. This implies that in the gadgets, eventually 
only $(q,\L)$ states are visited. Then, according to the definition
of $\straa_G$, the rank in prefixes of $\rho_G$ decreases and eventually
reaches rank $0$, that is a state with priority~$0$ is visited, and we 
have again a contradiction.
\end{itemize}
Therefore, $\straa_G$ is a winning strategy in the game~$G$.

Second, we show that if player~$1$ has a winning strategy $\straa_G$ in $G$,
then we can construct an almost-sure winning strategy $\straa_M$ in $M$.
By the result of~\cite[Lemma~8]{CD10a} and its proof, 
we can assume that $\straa_G$ is \emph{energy-based memoryless}, that is
$\straa_G(\rho) = \straa_G(\rho')$ for all $\rho, \rho'$ such that
$\Last(\rho) = \Last(\rho')$ and $\EL(\rho) = \EL(\rho')$.
In particular, if $h(\rho) = h(\rho')$, then $\straa_G(\rho) = \straa_G(\rho')$.

We define the strategy $\straa_M$ as follows: for each prefix $\rho_M$ in $M$,
let $\straa_M(\rho_M) = \straa_G(\rho)$ where $\rho$ is such that $h(\rho) = \rho_M$. 
By the above remark, the strategy~$\straa_M$ is uniquely and well defined.
We also know that $\straa_G$ uses finite memory. 
Therefore, in $G_{\straa_G}$ all cycles are have nonnegative energy
and visit a priority~$0$ state. Therefore, all cycles in $M_{\straa_M}$
have nonnegative energy; and if there is a reachable closed recurrent set~$U$ in $M_{\straa_M}$
that contains only priority~$1$ states, then in $G_{\straa_G}$ player~$2$ 
can fix a strategy to reach the closed recurrent set~$U$ (by choosing the successor
of probabilistic states using $(\cdot,\R)$ states) and in the states of~$U$, 
player~$2$ always chooses $(\cdot,\L)$ states. The (unique) outcome is a play 
that eventually remains in the closed recurrent set and therefore visits priority~$1$
states only from some point on, spoiling strategy $\straa_G$, a contradiction.
Hence, all closed recurrent sets in $M_{\straa_M}$ contain a priority~$0$ state
and the B\"uchi objective is satisfied with probability $1$ under strategy $\straa_M$.
\qed
\end{proof}


\begin{proof}[of Lemma~\ref{lem:energy-parity-MDP}]
Consider the construction of $\tuple{M',p',\weight'}$ defined before 
Lemma~\ref{lem:energy-parity-MDP}.
Let $\mathit{Win}' \subseteq  Q \times\set{0,2,\ldots, 2r}$ be the set of almost-sure 
winning states in $M'$ for the energy B\"uchi
objective and let $\mathit{Win} = \set{q \in Q \mid \exists i \cdot (q,2i) \in \mathit{Win}'}$
be the projection of $\mathit{Win}'$ on $Q$. 
We then convert all states in $W$ to absorbing (or sink) states with weight~0,
and then consider almost-sure energy B\"uchi winning set $Z$ with $\mathit{Win}$ as 
the B\"uchi set (this is almost-sure energy and reachability to $\mathit{Win}$).

We claim $Z$ is the almost-sure winning set for energy parity in $M$. 
The proof is as follows. Let $\ov{Z}=Q \setminus Z$. 
Consider an arbitrary strategy $\straa$ for player~1 and a starting state 
$q \in \ov{Z}$.
Assume towards contradiction that $\straa$ is almost-sure winning for energy 
parity objective.
Suppose there is an end-component $U$ such that $U \cap \ov{Z}\neq \emptyset$, 
that is visited infinitely often with positive probability. 
Since $\straa$ is almost-sure winning, we must have that $\min(p(U))$ is even 
(say $2i$) and the energy objective is satisfied. 
Hence in the copy $2i$ in $M'$ we have that $U$ is almost-sure winning. 
This means $U\times \set{2i} \subseteq \mathit{Win}'$ and since 
$U \subseteq Q$ we have $U \subseteq \mathit{Win}$.
But this contradicts that $U \cap \ov{Z} 
\neq \emptyset$ and $\mathit{Win} \subseteq Z$.
It follows that there is no end-component that intersects with $\ov{Z}$ 
that is visited infinitely often with positive probability.
Hence, given $\straa$, the set $Z$ must be reached with probability~1.
If the energy objective is also ensured with probability~1 by $\straa$, then 
the strategy is almost-sure winning for energy and reachability to $\mathit{Win}$ 
(since from $Z$ almost-sure winning for energy and reachability to $\mathit{Win}$ can
be ensured).
This shows that $q$ would belong to $Z$. 
This is a contradiction and completes the proof.
\qed
\end{proof}

\smallskip\noindent{\bf Bound for strategies.} We construct an almost-sure winning 
strategy of size at most $2 \cdot (|Z|+1) \cdot W$ as follows. 
We first partition the set $\mathit{Win}$ as follows: $\mathit{Win}_0$ is the set
of states that is winning in copy~0; $\mathit{Win}_2$ is the set of states that 
is winning in copy~2 and not in copy~0; and so on. 
For a state $q \in \mathit{Win}$, let $q \in \mathit{Win}_{2i}$, then for the state
$q$ we play the almost-sure winning strategy for in copy~$2i$. 
Since the copies are disjoint, the total memory required for the almost-sure winning 
strategies is 
$\sum_{i} 2\cdot |\mathit{Win}_{2i}| \cdot W = 2\cdot |\mathit{Win}| \cdot W$.
For states $q \in Z \setminus \mathit{Win}$, we play the almost-sure winning strategy 
to reach $\mathit{Win}$ ensuring the energy objective. 
Since for the reachability to $\mathit{Win}$ we can consider states in $\mathit{Win}$ 
as a single absorbing state, the memory required is at most 
$2\cdot (|Z \setminus \mathit{Win}|+1) \cdot W$.
After reaching $\mathit{Win}$ the strategy switches to the almost-sure winning strategy 
from $\mathit{Win}$.
Hence the total memory required by the almost-sure winning strategy is at most 
$2\cdot (|Z|+1)\cdot W$.

\smallskip\noindent{\bf Algorithm.} If we simply apply the algorithm for energy B\"uchi MDPs
on the reduction, then we obtain a $O(|E| \cdot d \cdot (d \cdot |Q|)^5 \cdot W)$ algorithm.
The improved version is obtain by simply following the steps of the proof. 
First, for each copy we compute the almost-sure winning set for the energy B\"uchi objective
Since each copy is disjoint and in each copy we require $O(|E|  \cdot |Q|^5 \cdot W)$,
the total time required to compute the $\mathit{Win}$ is at most 
$O(|E|  \cdot d\cdot |Q|^5 \cdot W)$.
Finally the almost-sure energy reachability to $\mathit{Win}$ can be achieved in an additional
$O(|E|  \cdot |Q|^5 \cdot W)$ time.
Hence we obtain an $O(|E|  \cdot d\cdot |Q|^5 \cdot W)$ time algorithm.


\section{Details of Section 4}

\begin{proof}[of Lemma~\ref{lemm:key}]
  The strategy $\straa^*$ for the mean-payoff parity objective
  is produced by combining two pure memoryless strategies: 
  $\straa_m$ for the expected mean-payoff objective and 
  $\straa_Q$ for the objective of reaching the smallest priority.
  We present a few properties that we use in 
  the correctness proof of the almost-sure winning strategy.
  \begin{enumerate}
  \item \emph{Property~1. Finite-time reach to smallest priority.} 
  Observe that under the strategy $\straa_Q$ we obtain a Markov chain
  such that every closed recurrent set in the Markov chain contains
  states with the smallest priority, and hence from all states~$q$ a
  state with the smallest priority (which is even) is reached in
  finite time with probability~1. 
  \item \emph{Property~2. Uniform value.} The expected mean-payoff value for all
  states $q \in Q$ is the same: if we fix the memoryless strategy $\straa_u$ 
  that chooses all successors uniformly at random, then we get a Markov 
  chain as the whole set $Q$ as a closed recurrent set, and hence from
  all states $q\in Q$ any state $q' \in Q$ is reached in finite time
  with probability~1, and hence the expected mean-payoff value at $q$ is at
  least the expected mean-payoff value at $q'$.  It follows that for all $q,q'\in Q$ 
  the expected mean-payoff value at $q$ and $q'$ coincide.
  Let us denote the uniform expected mean-payoff value by $v^*$. 

 \item \emph{Property~3. Property of optimal mean-payoff strategy.} 
  The strategy $\straa_m$ is a pure
  memoryless strategy and once it is fixed we obtain a Markov chain.
  The limit of the average frequency (or Cesaro limit) exists for all states 
  and since $\straa_m$ is optimal it follows that for all states $q \in Q$
  we have
  \[
  \lim_{n \to \infty} \frac{1}{n} \cdot \sum_{i=1}^n
  \Exp_q^{\straa_m}[\weight((\theta_i,\theta_{i+1}))] = v^*,
   \] 
  where $\theta_i$ is the random variable for the $i$-th state of a
  path. 
  In the resulting Markov chain obtained by fixing $\straa_m$, the expected 
  mean-payoff value for every closed recurrent set must be $v^*$; otherwise, 
  if there is a closed recurrent set with expected mean-payoff value less than 
  $v^*$, then there must be a closed recurrent set with expected mean-payoff 
  value greater than $v^*$ as all states have 
  the uniform value $v^*$, but then we obtain a state with expected mean-payoff 
  value greater than $v^*$ which contradicts Property~2.
  Hence from the theory of finite state Markov chains (the almost-sure 
  convergence to the Cesaro limit~\cite{Norris}) we obtain that 
  \[
  \Prb_{q}^{\straa}(\set{\rho \mid \lim_{\ell\to\infty}\frac{1}{\ell}\cdot \EL(\weight,\rho(\ell)) \geq v^*} ) =
  \lim_{\ell\to \infty} \Prb_{q}^{\straa}(\set{\rho \mid \frac{1}{\ell}\cdot \EL(\weight,\rho(\ell)) \geq v^*} ) = 1.
  \]
  In the above equality the limit and the probability operators are exchanged using 
  Lesbegue's Dominated Convergence Theorem~\cite{Royden88} (as the weights are bounded).
  Hence for all $\epsilon>0$, 
  there exists $j(\epsilon) \in \nats$ such that if $\straa_m$ is played for any 
  $\ell \geq j(\epsilon)$ steps then the average of the weights for 
  $\ell$ steps is at least $\epsilon$ within the expected mean-payoff value of the MDP with 
  probability at least $1-\epsilon$, 
  i.e., for all $q \in Q$, for all $\ell \geq j(\epsilon)$ we have
  \[
  \Prb_{q}^{\straa}(\set{\rho \mid \frac{1}{\ell}\cdot \EL(\weight,\rho(\ell)) \geq v^*-\epsilon} ) \geq 1-\epsilon.
   \]
 \end{enumerate}
  Let $W$ be the maximum absolute value of the weights.  The
  almost-sure strategy $\straa^*$  for mean-payoff parity objective is played in rounds, and
  the strategy for round $i$ is as follows:
  \begin{enumerate}
  \item \emph{Stage 1.} First play the strategy $\straa_Q$ till the
    smallest priority is reached.
  \item \emph{Stage 2.} Let $\epsilon_i=1/i$.  If the game was in the
    first stage in this ($i$-th round) for $k_i$ steps, then play the
    strategy $\straa_m$ for $\ell_i$ steps such that $\ell_i \geq
    \max\set{j(\epsilon_i), i \cdot k_i\cdot W}$.  This ensures
    that the with probability at least $1-\epsilon_i$ the average of the 
    weights in round $i$ is at least
    \[
    \begin{array}{rcl}
      \displaystyle
      \frac{\ell_i\cdot (v^* -\epsilon_i) - k_i \cdot W}{k_i + \ell_i} & = & 
      \displaystyle 
      \frac{(\ell_i+k_i)\cdot v^*- \ell_i \cdot \epsilon_i -k_i\cdot v^* - k_i\cdot W} 
      {k_i + \ell_i} \\[2ex]
      & \geq & 
     \displaystyle 
      v^*- \frac{\ell_i \cdot \epsilon_i + k_i\cdot v^* + k_i \cdot W}{\ell_i+k_i} \\[2ex]
     & \geq & 
      \displaystyle 
      v^*-\epsilon_i - \frac{2 \cdot k_i\cdot W}{\ell_i+k_i} 
      \quad (\text{since } v^* \leq W)\\[2ex]
      & \geq & 
      \displaystyle 
      v^* - \epsilon - \frac{2\cdot k_i \cdot W}{\ell_i}\\[2ex]
      & \geq & 
      \displaystyle 
      v^* -\epsilon_i - \frac{2}{i} \quad (\text{since } \ell_i \geq i \cdot k \cdot W) \\[2ex]
      & = & 
      \displaystyle 
      v^* - \frac{3}{i}.
    \end{array}
    \]
   Then the strategy proceeds to round $i+1$.
  \end{enumerate}
  The strategy ensures that there are infinitely many
  rounds (this follows by Property~1 of finite-time reachability to min even 
  priority state). Hence with probability~1 the smallest priority that is visited 
  infinitely often is the smallest priority of the end-component (which is even).  
  This ensures that the parity objective is satisfied with probability~1.  
  We now argue that the mean-payoff objective is also satisfied with 
  probability~1.
  Fix arbitrary $\epsilon >0$ and consider $i$ such that $\frac{3}{i} \leq \epsilon$.
  For all $j \geq i$, in round $j$, the average weights is at least $v^* -\epsilon$ 
  with probability at least $1-\epsilon$. 
  Since mean-payoff objective is independent of finite prefixes, 
  for all $q \in Q$ we have 
  \[
  \Prb_{q}^{\straa}(\set{\rho \mid \lim_{\ell \to \infty} 
  \frac{1}{\ell}\cdot \EL(\weight,\rho(\ell)) \geq v^*-\epsilon} ) \geq 1-\epsilon.
  \] 
  Since $\epsilon>0$ is arbitrary, letting $\epsilon \to 0$, we obtain that 
  for all $q \in Q$ we have 
  \[
   \Prb_{q}^{\straa}(\set{\rho \mid \lim_{\ell \to \infty} 
  \frac{1}{\ell}\cdot \EL(\weight,\rho(\ell)) \geq v^*} ) \geq 1
  \]
  Hence depending on whether $v^* \geq \nu$ or $v^* > \nu$ we obtain 
  the desired result.
  \qed
\end{proof}

\smallskip\noindent{\bf Further details about computing winning end-components for MDPs with 
mean-payoff parity objectives.} 
We now present some further details about computing the winning end-components with 
mean-payoff parity objectives. 
The computation of the winning end-components is done iteratively 
by computing winning end-components with smallest priority~0, 
then winning end-components with smallest priority~2, and so on.
We start with the initial MDP $M_0:=M$. 
In iteration $i$ the remaining MDP is $M_i$. 
We compute the maximal end-component decomposition of $M_i$,
then consider the maximal-end components $U$ that contains 
only states with priority at least $2i$, and at least one state with 
priority $2i$. 
If there is such an end component $U$ where the expected 
mean-payoff value is at least $\nu$ at some state, then $U$ 
is included in $W_{2i}$. 
The we consider the \emph{random attractor} (i.e., alternating 
reachability to $W_{2i}$ by the random player) to $W_{2i}$ and 
the set of random attractor is removed from the MDP for the next
iteration.
The random attractor to a set $T$ is as follows: $T_0:=T$ and 
for $i \geq 0$ we have $T_{i+1} := T_i \cup \set{q \in Q_1 \mid \forall q' \in Q. 
(q,q') \in E \rightarrow q' \in T_i} \cup \set{q \in Q_P \mid \exists q' \in Q. 
(q,q') \in E \land q' \in T_i}$ and the random attractor is 
$\bigcup_{i \geq 0} T_i$.
It follows from the results of~\cite{CH11} (see Lemma 2.1 of~\cite{CH11})
that if we consider a set of end-components, and take random attractor
to the set, then the maximal end-component decomposition of the remaining
MDP remains unaffected. 
Moreover, the complement of a random attractor in an MDP is always a
sub-MDP.
The set $W = \bigcup_{i=0}^{\lfloor d/2 \rfloor} W_{2i}$ is the union 
of the states of the winning end-components. 
The formal pseudocode is given as Algorithm~\ref{algo:winendec}.

\begin{algorithm}[t]
\caption{\bf AlgoWinEndComponent}
\label{algo:winendec}
{ 
\begin{tabbing}
aa \= aa \= aa \= aa \= aa \= aa \= aa \= aa \= aa \= aa \= aa \= aa \kill
\> \textbf{Input}: An MDP $M$ with parity function $p$, weight function $\weight$ and threshold $\nu$. \\
\> \textbf{Output}: The set $W$ of union of winning end-components. \\ 
\> 1. $i := 0$; \\
\> 2. $M_{0} := M$;  \\
\> 3. \textbf{For} $i:=0$ to ${\lfloor d/2 \rfloor}$ {\bf do} \\
\>\> 3.1 Compute the maximal end-component decomposition of $M_i$; \\
\>\> 3.2 Let $\calw_{2i}$ be the maximal end-components $U$ \\
\>\>\> such that 
    $U \subseteq \bigcup_{j\geq 2i} p^{-1}(j)$ and $U \cap p^{-1}(2i) \neq \emptyset$; \\
\>\> 3.3 Let $\calw'_{2i} \subseteq \calw_{2i}$ be the set of maximal end-components $U \in \calw_{2i}$ such that  \\
\>\>\> in the sub-MDP induced by to $U$ there exists $q$ with $\va(\MeanPayoff(w)) \geq \nu$. \\
\>\> 3.4 $W_{2i}:= \bigcup_{U \in \calw'_{2i}} U$; \\
\>\> 3.5 $Z_{2i}:=$ Random attractor of $W_{2i}$ in $M_i$; \\
\>\> 3.6 $M_{i+1}:=$ sub-MDP induced by removing $Z_{2i}$ in $M_i$; \\
\>\> 3.7 $i:= i+1$; \\

\> 4. \textbf{return} $W := \bigcup_{i=0}^{\lfloor d/2 \rfloor} W_{2i}$.
\end{tabbing}
}
\end{algorithm}

\section{Details of Theorem~\ref{thrm:disjunction}}\label{sec:other}
We first consider the disjunction of mean-payoff and parity objectives, and 
then disjunction of energy and parity objectives.

\smallskip\noindent{\bf Disjunction of mean-payoff and parity objectives.} 
For the disjunction of mean-payoff and parity objectives in MDPs we consider 
end-components analysis. An end-component is winning if either the parity objective
can be ensured almost-surely, or the mean-payoff objective can be ensured 
almost-surely. 
Since determining almost-sure winning for parity objective and mean-payoff 
objectives can be done in polynomial time, we can use the algorithm of 
Section~\ref{sec:MDP-mean-payoff-parity} for computing winning end-components 
and then reachability to winning end-components. 
Hence disjunction of mean-payoff parity objectives can be solved in polynomial
time, and also pure memoryless optimal strategies exist.

\smallskip\noindent{\bf Disjunction of energy and parity objectives.} 
The solution of disjunction of energy and parity objectives is achieved using
the end-component analysis: an end-component is winning if either the parity
objective can be ensured almost-surely or the energy objective can be ensured
almost-surely.
Whether an end-component is almost-sure winning for parity can be decided
in polynomial time, and for energy objectives it is in NP $\cap$ coNP. 
Hence the winning end-components can be determined in NP $\cap$ coNP.
Let $W_1$ be the union of the set of winning end-components for almost-sure 
parity, and let $W_2$ be the union of the set of remaining winning 
end-components (i.e., only almost-sure winning for energy). 
Finally we need to ensure almost-sure reachability to $W_1$ or almost-sure 
energy reachability to $W_2$. 
Again this can be achieved in NP $\cap$ coNP.

The desired result follows (also see discussion section in~\cite{CD11-TechRpt}).

\end{document}